\documentclass[nologo,url,11pt,a4paper]{ETHpaper}
\usepackage{amsmath,amssymb,amsthm,subfigure,hyperref}
\usepackage[english]{babel}
\usepackage{graphicx}
\usepackage{siunitx}
\usepackage[numbers,sort&compress]{natbib}

\usepackage[table]{xcolor}
\newcommand{\todo}[1]{}

\title{Predicting Scientific Success Based on Coauthorship Networks}

\author{Emre Sarig\"ol, Ren\'{e} Pfitzner$^*$, Ingo Scholtes, Antonios Garas, Frank Schweitzer}

\address{Chair of Systems Design, ETH Zurich \\
CH-8092 Zurich, Switzerland \\
$^*$\texttt{rpfitzner@ethz.ch}}

\www{\url{http://www.sg.ethz.ch}}

\newcommand{\mean}[1]{\left\langle #1 \right\rangle}

\newcolumntype{.}{D{.}{.}{-1}}

\begin{document}

\maketitle

\begin{abstract}
We address the question to what extent the success of scientific articles is due to social influence.
Analyzing a data set of over \num{100000} publications from the field of Computer Science, we study how centrality in the coauthorship network differs between authors who have highly cited papers and those who do not.
We further show that a machine learning classifier, based only on coauthorship network centrality measures at time of publication, is able to predict with high precision whether an article will be highly cited five years after publication.
By this we provide quantitative insight into the social dimension of scientific publishing -- challenging the perception of citations as an objective, socially unbiased measure of scientific success.
\end{abstract}

\section{Introduction}
\label{sec:intro}
%
Quantitative measures are increasingly used to evaluate the performance of research institutions, departments, and individual scientists.
Measures like the absolute or relative number of published research articles are frequently applied to quantify the \emph{productivity} of scientists.
To measure the \emph{impact} of research, citation-based measures like the total number of citations, the number of citations per published article or the h-index~\cite{Hirsch2005}, have been proposed.
Proponents of such citation-based measures or rankings argue that they allow to quantitatively and objectively assess the \emph{quality} of research, thus encouraging their use as simple proxies for the \emph{success} of scientists, institutions or even whole research fields.
The intriguing idea that by means of citation metrics the task of assessing research quality can be ``outsourced'' to the \emph{collective intelligence} of the scientific community, has resulted in citation-based measures becoming increasingly popular among research administrations and governmental decision makers.
As a result, such measures are used as one criterion in the evaluation of grant proposals and research institutes or in hiring committees for faculty positions.
Considering the potential impact for the careers of - especially young - scientists, it is reasonable to take a step back and ask a simple question: To what extent do \emph{social factors} influence the number of citations of their articles?
Arguably, this question challenges the perception of science as a systematic pursuit for objective truth, which ideally should be free of personal beliefs, biases or social influence.
On the other hand, quoting Werner Heisenberg~\cite{heisenberg1969}, \emph{``science is done by humans''}, it would be surprising if specifically scientific activities were free from the influences of social aspects.
Whereas often the term ``social influence'' has a negative connotation, we don't think that social influence in science
necessarily stems from malicious or unethical behavior, like e.g. nepotism, prejudicial judgments, discrimination or in-group favoritism.
We rather suspect that, as a response to the increasing amount of published research articles and our limited ability to keep track of potentially relevant works, a growing importance of social factors in citation behavior is due to natural mechanisms of \emph{social cognition} and \emph{social information filtering}.

In this paper we address this issue by studying the influence of social structures on scholarly citation behavior.
Using a data set comprising more than \num{100000} scholarly publications by more than \num{160000} authors, we extract time-evolving coauthorship networks and utilize them as a proxy for the evolving social network of the scientific discipline \emph{computer science}.
Based on the assumption that the centrality of scientists in the resulting social network is indicative for the \emph{visibility} of their work, we then study to what extent the \emph{``success''} of research articles in terms of citations can be predicted using only knowledge about the embedding of authors in the social network \textit{at time of publication}.
Our prediction method is based on a random forest classifier and utilizes a set of complementary network centrality measures.
We find strong evidence for our hypothesis that authors whose papers are highly cited in the future have - on average - a significantly higher centrality in the social network at the time of publication.
Remarkably, we are able to predict whether an article will belong to the $10 \%$ most cited articles with a precision of $60 \%$.
We argue that this result quantifies the existence of a \textit{social bias}, manifesting itself in terms of visibility and attention, and influencing measurable citation ``success'' of researchers.
The presence of such a social bias not only highlights problems with current publication and citation practices.
It also threatens the interpretation of citations as \emph{objectively awarded esteem}, which is the justification for using citation-based measures as universal proxies of \emph{quality} and \emph{success}.

The remainder of this article is structured as follows:
In section~\ref{sec:citations} we review a number of works that have studied scientific collaboration structures as well as their relation to citation behavior.
In section~\ref{sec:methods} we describe our data set and provide details of how we construct time-evolving coauthorship networks.
We further introduce a set of network-theoretical measures which we utilize to quantitatively assess the centrality and embedding of authors in the evolving coauthorship network.
In section~\ref{sec:hypotheses} we introduce a number of hypotheses about the relations between the position of authors in the coauthorship network and the future success of their publications.
We test these hypotheses and obtain a set of candidate measures which are the basis for our prediction method described in section~\ref{sec:prediction}.
We summarize and interpret our findings in section~\ref{sec:conclusion} and discuss their implications for the application of citation-based measures in the quantitative assessment of research.

\section{The Complex Character of Citations}
\label{sec:citations}

It is remarkable that, even though citation-based measures have been used to quantify research impact since almost sixty years~\cite{Garfield1955}, a complete \textit{theory of citations} is still missing.
In particular, researchers studying the social processes of science have long been arguing that citations have different, complex functions that go well beyond a mere attribution of credit~\cite{Leydesdorff1998}.
At the level of scientific articles, a citation can be interpreted as a ``discursive relation'', while at the level of authors citations have an additional meaning as expression of ``professional relations''~\cite{Leydesdorff1998}.
Additional interpretations have been identified at aggregate levels, like e.g. social groups, institutions, scientific communities or even countries citing each other.
These findings suggest that citations are indeed a complex phenomenon which have both cognitive and a social dimension~\cite{Leydesdorff1998,Nicolaisen2003}.
This questions an oversimplified interpretation of citations as objective quality indicator.
The complex character of scholarly citations was further emphasized recently~\cite{Laloe2009}.
Here, the authors argue that, apart from an attribution of scientific merit, references in scientific literature often serve as a tool to guide and orient the reader, to simplify scientific writing and to associate the work with a particular scientific community.
Furthermore, they highlight that citation numbers of articles are crucially influenced not only by the popularity of a research topic and the size of the scientific community, but also by the number of authors as well as their prominence and visibility.

Facilitated by the wide-spread availability of scholarly citation databases, some advances in the understanding of the dynamics of citations have been made in the last years.
Generally, citation practices seem to differ significantly across different scientific disciplines, 
and thus complicating the definition of universal citation-based impact measures.
However, the remarkable finding that -- independent of discipline -- citations follow a log-normal distribution and can be rescaled in such a way that citation numbers become comparable~\cite{RadForCas2008,Stringer2010}, suggests that the mechanisms behind citation practices are universal across disciplines, and differences are mainly due to differing community sizes.

Additionally to investigations of the differences across scientific communities, the relations between citations and coauthorships were studied in recent works.
Using data from a number of scientific journals, it was shown that the citation count of an article is correlated both with the number of authors and the number of institutions involved in its production~\cite{Katz1997,Figg2006}.
Studying data from eight highly ranked scientific journals, it was shown~\cite{HsuHua} that a) single author publications consistently received the lowest number of citations and b) publications with less than five coauthors received less citations than the average article.
Studying citations between individuals rather than articles, in~\cite{MarBalKarNew2013} it was observed that coauthors tend to cite each other sooner after the publication of a paper (compared to non-coauthors).
Further, the authors showed that a strong tendency towards reciprocal citation patterns exists.
Although these findings already indicate that social aspects influence citing behavior, in this work we are going to quantitatively reveal the true extent of this influence.

Going beyond a mere study of direct coauthorship relations, first attempts to study \emph{both} citation and coauthorship structures from a \emph{network perspective} have been made recently.
Aiming at a measure that captures both the \emph{amount} as well as the \emph{reach} of citations in a scientific community, a citation index that incorporates the distance of citing authors in the collaboration network was proposed~\cite{BrasAmoros2011}.
Another recent study~\cite{Wallace2012} used the topological distance between citing authors in the coauthorship network to extend the notion of self-citations.
Interestingly, apart from direct self-citations, this study could not find a strong tendency to cite authors that are close in the coauthorship network.

Different from previous works, in this article we study correlations between the \emph{centrality} of authors in collaboration networks and the \textit{citation success} of their research articles.
By this we particularly extend previous works that use a network perspective on coauthorship structures and citation patterns.
Stressing the fact that \textit{social relations} of authors play an important role for how much attention and recognition their research receives, we further contribute a quantitative view on previously hypothesized relations between the \emph{visibility} of authors and citation patterns.

\section{Time-Evolving Collaboration and Citation Networks}
\label{sec:methods}

In this work we analyze a data set of scholarly citations and collaborations obtained from the Microsoft Academic Search~\footnote{http://academic.research.microsoft.com} (MSAS) service.
The MSAS is a scholarly database containing more than \num{35} Million publication records from \num{15} scientific disciplines.
Using the Application Programming Interface (API) of this service, we extracted a subset of more than \num{100000} computer science articles, published between 1996 and 2008, in the following way:
First, we retrieved unique numerical identifiers (IDs) of the \num{20000} highest ranked authors in the field of \emph{computer science}.
This ranking is the result of an MSAS internal ``field rating'', taking into account several scholarly metrics of an author (number of publications, citations, h-index) and comparing them to the typical values of these metrics within a certain research field.
As the goal was to build coauthorship and citation networks of reasonable size, in a second step we chose \num{1000} authors i.i.d. uniformly from the set of these \num{20000} authors.
In the third step, we obtained information on coauthors, publication date, as well as the list and publication date of citing works for all the publications authored by these \num{1000} authors between 1996 and 2008.
This results in a data set consisting of a total of \num{108758} publications  from the field computer science, coauthored by a total of \num{160891} researchers.
Each publication record contains a list of author IDs, which, by means of disambiguation heuristics internally applied by the MSAS service, uniquely identify authors independent of name spelling variations.
The absence of name ambiguities is one feature that sets this data set apart from other data sets on scholarly publications that are used frequently.
Based on this data set we extracted a \emph{coauthorship network}, where nodes represent authors and links represent coauthorship relations between authors.
In addition, using the information about citing papers, we extracted \emph{citation dynamics}, i.e. the time evolution of the number of citations of all publications in our data set.
Similar to earlier works, we argue that the coauthorship network can be considered a first-order approximation of the complete scientific collaboration network~\cite{MarBalKarNew2013}.
Based on the publication date of an article, we additionally assign time stamps to the extracted coauthor links -- thus obtaining time-evolving coauthorship networks.

We analyze the evolution of the coauthorship network using a sliding window of two years in which we aggregate all coauthorships occurring within that time.
Starting with 1996, we slide this window in one year increments and obtain a total of 11 time slices representing the evolution of collaboration structures between 1996 and 2008.
We use an extended time-window of two years to account for the continuing effect of a coauthorship in terms of awareness about the coauthors works.
Although larger time windows are certainly possible (and their effects interesting to investigate), in this work we are less concerned with the optimal time-window size and consistently use the above described approach.
However, performed consistency checks with varying time-window sizes suggest robustness of our results.

Table~\ref{table:NetworkSizes} summarizes the number of nodes and links in the coauthorship network, the number of publications in each time slice as well as the fractional size of the largest connected component (LCC).
Note that the time-aggregated network (overall) forms one giant component with only a minor fraction of isolated nodes, whereas some of the time slices fall apart into many separated components.
Note also that the size of the largest connected component is increasing with time, which may indicate either a possible bias in the coverage of the MSAS database to favour newer articles, or an increase of ``collaborativeness'' in science.
As we are going to perform a social network analysis of the collaboration time slices -- and some measures (like eigenvector centrality) are not well-defined for unconnected graphs -- we apply all of the following analysis always on the largest connected component.
For each network corresponding to one two-year time slice, we compute a number of node-level metrics that allow us to quantitatively monitor the evolution of network positions for all authors.
In particular, we compute \emph{degree centrality}, \emph{eigenvector centrality}, \emph{betweenness centrality} and \emph{$k$-core centrality} of authors.
For details on the used centrality measures, please refer to the Supplementary Material or the textbook by Newman~\cite{NewmanBook}.
Here we utilize implementations of these measures provided by the igraph package~\cite{igraph}.
\begin{table}
\centering
\begin{tabular}{|
c
c
S[table-number-alignment=center, table-format = 1.2]
S[table-number-alignment=center, table-format = 8.0]
S[table-number-alignment=center, table-format = 6.0]
S[table-number-alignment=center, table-format = 6.0]
|
}
    \hline
    & {Year} & {LCC fraction} & {Links} & {Nodes} &  {Publications}  \\
    \hline
  	& 1996-1997 & 0.18 & 61046 & 2845 & 1160 \\
    \hline	
  	& 1997-1998 & 0.37 & 130938 & 6381 & 3070 \\
    \hline	
  	& 1998-1999 & 0.45 & 153412 & 8470 & 4054 \\
    \hline
  	& 1999-2000 & 0.50 & 186318 & 10413 & 5320 \\
    \hline
  	& 2000-2001 & 0.60 & 358188 & 13451 & 6561 \\
    \hline
  	& 2001-2002 & 0.63 & 413846 & 15309 & 7026 \\
    \hline
  	& 2002-2003 & 0.74 & 542912 & 20238 & 9193 \\
    \hline
  	& 2003-2004 & 0.77 & 653224 & 23624 & 10608 \\
    \hline
  	& 2004-2005 & 0.79 & 745352 & 26258 & 11430 \\
    \hline
  	& 2005-2006 & 0.83 & 889996 & 29886 & 12919 \\
    \hline
  	& 2006-2007 & 0.84 & 914614 & 32412 & 13568 \\
    \hline
  	& 2007-2008 & 0.86 & 858554 & 35255 & 14214 \\
    \hline
	& Overall & 0.99 & 5324330 & 160891 & 108758 \\
    \hline

  \end{tabular}
  \caption {Number of papers and size of the collaboration network 2-year subgraphs between 1995-2008 used in our study.}
  \label{table:NetworkSizes}
\end{table}

A major focus of our work is to assess the predictive power of an author's position in the coauthorship network for the citation success of her future articles.
To do so we adopt a so called \emph{hindcasting approach}:
For each publication $p$ published in a given year $t$, we extract the list of coauthors as well as the LCC of the coauthorship network in the time slice $[t-2,t]$, and calculate the centrality measures.
Based on the citation data, we furthermore calculate the number of citations $c_p$ paper $p$ gained within a time frame of \textit{five years} after publication, i.e. in the time slice $[t, t+5]$.

In particular, we are interested in those publications that are among the most successful ones.
Defining \textit{success} is generally an ambiguous endeavor.
As justified in the introduction, here we take the (controversial) viewpoint that success is directly measurable in number of citations.
We specifically focus on a very simple notion of success in terms of \emph{highly cited papers} and, similar to~\cite{Newman2009}, assume that a paper is \textit{successful} if five years after publication it has more citations than $90 \%$ of all papers published in the same year.
We refer to the set of successful papers in year $t$ as $P_{\uparrow}(t)$.
The set of remaining papers, i.e. those published at time $t$ that are cited less frequently than the top $10 \%$, is denoted as $P_{\downarrow}(t)$.
\pagebreak
\section{Statistical Dependence of Coauthorship Structures and Citations}
\label{sec:hypotheses}

Having a large social network and ``knowing the right people'' often is a prerequisite for career success.
However, science is often thought to be one of the few fields of human endeavor where success depends on the quality of an authors' work, rather than on her social connectedness.
Given the time evolving coauthorship network, as well as the observed success (or lack thereof) of a publication, we investigate two research questions, aiming to quantify the aspect of social influence on citation success.
First, we examine whether there is a general statistical dependency of central authors in the coauthorship network to publish papers that are more successful than non-central.
Second, we investigate whether the inverse effect is present and the success of a paper influences the future coauthorship centrality of its authors.
\subsection{Effects of Author Centrality on Citation Success}
\label{sec:hypotheses:H1}
To quantify the first research question we test the following hypothesis.

{ \textbf{H1:} \em At the time of publication, authors of papers in $P_{\uparrow}(t)$ are more central in the coauthorship network than authors of articles in $P_{\downarrow}(t)$.}

As papers often have more than one author, for each paper we will consider only the coauthorship network centralities of the author with the highest coauthorship degree, and refer to this as the \textit{coauthorship centrality of the paper}.
This choice is motivated by the intuition that the centrality of the best connected coauthor should provide the major amount of (socially triggered) visibility for the publication.
We test \textbf{H1} by comparing coauthorship centrality distributions of papers in $P_{\uparrow}(t)$ and $P_{\downarrow}(t)$ for each year $t$.
In order to compare the centrality distributions, we apply a \emph{Wilcoxon-Mann-Whitney two-sample rank-sum-test}~\cite{Mann1947}.
For each of the four centrality metrics we test the null hypothesis that coauthorship centrality distributions of papers in $P_{\uparrow}(t)$ and $P_{\downarrow}(t)$ are the same against the alternative hypothesis that the centrality distribution of papers in $P_{\uparrow}(t)$ is stochastically larger than the one of papers in $P_{\downarrow}(t)$.
The p-values of the tests as well as the corresponding averages and variances of the four considered centrality metrics in the two sets are shown in Table~\ref{table:WilcoxTestsH1}.
\begin{table}
\centering
\begin{tabular}{
|
c|
S[detect-weight, table-number-alignment=center, table-format = 1.2e-3]||
S[table-number-alignment=center, table-format = 1.2e-1]|
S[table-number-alignment=center, table-format = 1.2e-1]|
S[table-number-alignment=center, table-format = 1.2e-1]|
S[table-number-alignment=center, table-format = 1.2e-2]
|
}
      \hline
	& {\textbf{p-value}} 	& {$\mean {P_{\downarrow}}$} & {$\mean {P_{\uparrow}}$} & {var ${P_{\downarrow}}$}  & {var ${P_{\uparrow}}$}  \\
	\hline
	\textbf{$k$-core} & 1.28e-115 & 3.33e1 & 4.39e1 	& 1.20e4 & 7.18e3 \\
		\hline
		
	\textbf{Eigenvector} & 2.52e-34 & 2.87e-3 & 5.60e-4 			& 2.58e-3 & 5.40e-4 \\
		\hline
		
	\textbf{Betweenness} & 1.19e-68 & 6.30e5 & 1.52e6 & 4.19e12 & 1.58e13 \\
    	\hline
    	
	\textbf{Degree} & 5.63e-125 & 9.38e1 & 1.57e2 & 1.02e5 & 1.13e5 \\
    	\hline

\end{tabular}
\caption { P-values of one sided Wilcoxon-Mann-Whitney test. This quantifies whether the distribution of centralities of authors of articles in $P_{\uparrow}$ are (in a statistical sense) larger than those of authors of articles in $P_{\downarrow}$. Also shown are the means and variances of the centrality metrics in the two sets.}
\label{table:WilcoxTestsH1}
\end{table}
For all considered centrality metrics p-values are well below a significance level of 0.01.
This leads us to safely reject the null hypothesis, concluding that coauthorship centrality metrics of papers in $P_{\uparrow}(t)$ are stochastically larger than of those papers in $P_{\downarrow}(t)$.
This result indicates that all considered centrality metrics in the coauthorship network, at the time of publication of a paper, are indicative for future paper success.
Note however, that this statistical dependency is more complicated than the linear Pearson or the more general Spearman correlation.
Indeed, all the considered social network metrics are only weekly, if at all, correlated with citation numbers (see Supplementary Material).
To what extent citation success and coauthorship network centrality are statistically dependent is summarized in Table~\ref{table:TopAuthorIntersections}.
\begin{table}[t]
\centering
\begin{tabular}{|c|cccc|}
      \hline
	& \multicolumn{1}{c|}{Top $10 \%$ }
  & \multicolumn{1}{c|}{ Top $5 \%$ }
  & \multicolumn{1}{c|}{ Top $2 \%$ }
  & \multicolumn{1}{c|}{ Top $1 \%$ }  \\
	\hline
	\textbf{$k$-core} 	
	& \multicolumn{1}{c|}{\num{0.22}|\num{0.21}}
	& \multicolumn{1}{c|}{\num{0.17}|\num{0.16}}
	& \multicolumn{1}{c|}{\num{0.07}|\num{0.07}}
	& \multicolumn{1}{c|}{\num{0.01}|\num{0.01}} \\
		\hline
	\textbf{Eigenvector}
	& \multicolumn{1}{c|}{\num{0.11}|\num{0.11}}
	& \multicolumn{1}{c|}{\num{0.06}|\num{0.06}}
	& \multicolumn{1}{c|}{\num{0.01}|\num{0.01}}
	& \multicolumn{1}{c|}{\num{0.01}|\num{0.01}} \\
		\hline
	\textbf{Betweenness}
	& \multicolumn{1}{c|}{\num{0.20}|\num{0.20}}
	& \multicolumn{1}{c|}{\num{0.13}|\num{0.13}}
	& \multicolumn{1}{c|}{\num{0.11}|\num{0.11}}
	& \multicolumn{1}{c|}{\num{0.11}|\num{0.11}} \\
    	\hline
	\textbf{Degree}
	& \multicolumn{1}{c|}{\num{0.20}|\num{0.20}}
	& \multicolumn{1}{c|}{\num{0.15}|\num{0.15}}
	& \multicolumn{1}{c|}{\num{0.10}|\num{0.09}}
	& \multicolumn{1}{c|}{\num{0.07}|\num{0.07}} \\
    	\hline
	\textbf{Intersection}
	& \multicolumn{1}{c|}{\num{0.36}|\num{0.15}}
	& \multicolumn{1}{c|}{\num{0.27}|\num{0.11}}
	& \multicolumn{1}{c|}{\num{0.17}|\num{0.06}}
	& \multicolumn{1}{c|}{\num{0.12}|\num{0.04}} \\
    	\hline
        \hline
	\textbf{$\#$ papers}
	& \multicolumn{1}{c|}{\num{3700}}
	& \multicolumn{1}{c|}{\num{1844}}
	& \multicolumn{1}{c|}{\num{730}}
	& \multicolumn{1}{c|}{\num{362}} \\
    	\hline
\end{tabular}
\caption {First table entry indicates what fraction of papers, that have authors which are within the set of author with Top $x\%$ centrality metrics, are also Top $x\%$ of all papers in terms of citation success  ($P(\mathsf{toppaper}|\mathsf{topmetric})$).
Second table entry indicates what fraction of papers, that are Top $x\%$ of all papers in terms of citation success, have authors which are within the set of author with Top $x\%$ centrality metrics ($P(\mathsf{topmetric}|\mathsf{toppaper})$).
Row \textit{Intersection} indicates the intersection of all the above considered centrality metrics.}
\label{table:TopAuthorIntersections}
\end{table}
Left entry of each cell indicates what fraction of papers, that have authors with Top $x\%$ centrality metrics, belong to the Top $x\%$ of all papers in terms of citation success  ($P(\mathsf{toppaper}|\mathsf{topmetric})$).
Right entry of each cell indicates what fraction of papers, that are Top $x\%$ of all papers in terms of citation success, have authors which are within the set of authors with Top $x\%$ centrality metrics ($P(\mathsf{topmetric}|\mathsf{toppaper})$).
From these results, we conclude two observations:
First, the probabilities in every cell are well below $1$, indicating the absence of a simple linear (Pearson) correlation.
Second, especially considering $k$-core centrality, knowing a paper is Top $10\%$ successful, the conditional probability that it was written by an author with Top $10\%$ $k$-core centrality, is $P(\mathsf{topmetric}|\mathsf{toppaper})=0.21$.
Additionally, Table~\ref{table:TopAuthorIntersections} indicates that vice versa  $P(\mathsf{toppaper}|\mathsf{topmetric})=0.22$ of all papers that are published by authors with Top $10\%$ $k$-core centrality, are successful.
Considering the intersection of all four centrality metrics, we even find that $P(\mathsf{toppaper}|\mathsf{topmetric})=0.36$ of all papers published by the Top $10\%$ central authors in all four coauthorship centrality metrics, are within the Top $10\%$ cited papers.
We will use this observation as basis for a naive Bayes classifier in section~\ref{sec:prediction}.
\subsection{Coevolution of Coauthorship and Citation Success}
\label{sec:hypotheses:H23}
In the previous section we studied the question whether the centrality of authors in the coauthorship network is indicative for the success of publications in terms of citations.
Our results suggested that centrality in coauthorship networks is indeed indicative for citation success.
In the following we study the inverse relation and ask whether a shift in citation success of an author is indicative for her future position in the coauthorship network.
To answer this question, we consider all authors who published an article both at time $t$ and five years later at $t+5$.
We then categorize them based on the citation success of their articles published at time $t$ and time $t+5$.
We introduce two sets of authors:
Set $A_{\searrow}(t)$, which is the set of those authors who at time $t$ had at least one publication in class $P_{\uparrow}(t)$, but who at time $t+5$ did not have an article in class $P_{\uparrow}(t+5)$ anymore.
Set $A_{\nearrow}(t)$ contains all authors who at time $t$ had no article in class $P_{\uparrow}(t)$ but who at time $t+5$ published at least one article that falls in class $P_{\uparrow}(t+5)$.
In addition, we again record the coauthorship centralities of authors in these two sets for time windows $[t-2,2]$ and $[t+3,t+5]$.

For authors of set $A_{\nearrow}$ we test the following hypothesis:

{\textbf{H2:} \em Authors that experience a positive shift in their citation success (i.e. authors in $A_{\nearrow}$) will become more central in the coauthorship network.}

Complementary to {\textbf H2 }, for authors in set $A_{\searrow}$ we hypothesize:

{ \textbf{H3:} \em Authors that experience a negative shift in their citation success (i.e. authors in $A_{\searrow}$) will become less central in the coauthorship network.}

In order to test for \textbf{H2} and \textbf{H3}, we apply a \textit{pairwise Wilcoxon-Mann-Whitney} test.
To verify \textbf{H2} we test if the centralities of authors have decreased in the case of a decrease in publication success from time \textit{t} to \textit{t}+5.
To verify \textbf{H3} we test if the centralities of authors have increased in the case of an increase in publication success from time \textit{t} to \textit{t}+5.
\begin{table}
\centering
 \scalebox{0.8}{
\begin{tabular}{
|
c|
S[table-number-alignment=center,table-format=1.2e-2]
S[table-number-alignment=center,table-format=1.2e-2]
|
}
    \hline
centrality measure \& alternative & {$A_{\searrow}$} & {$A_{\nearrow}$}  \\
    \hline

    \textbf{$k$-core(t) > $k$-core(t+5)} & 3.15e-11 & 1 \\
    \hline

    \textbf{$k$-core(t) < $k$-core(t+5)} & 1 & 3.04e-55 \\
    \hline

    \textbf{ev-centr(t) > ev-centr(t+5)} & 5.18e-14 & 0.86  \\
    \hline

    \textbf{ev-centr(t) < ev-centr(t+5)} & 1 & 0.14 \\
    \hline

    \textbf{bw-centr(t) > bw-centr(t+5)} & 0.23 & 1 \\
    \hline

    \textbf{bw-centr(t) < bw-centr(t+5)} & 0.77	& 7.29e-30 \\
    \hline

    \textbf{degree(t) > degree(t+5)} & 6.69e-11 & 1  \\
    \hline

    \textbf{degree(t) < degree(t+5)} & 1 & 7.72e-62 \\
    \hline

     \textbf{$\#$ authors} & {521}  & {648} \\
     \hline
\end{tabular}
}
\caption { P-values of Wilcoxon-Mann-Whitney test for different coauthorship centralities and alternative hypotheses. Column $A_{\searrow}$ presents p-values for authors in set $A_{\searrow}$, column $A_{\nearrow}$ presents p-values for authors in set $A_{\nearrow}$.}
\label{table:WilcoxTestsH2H3}
\end{table}
Results of these hypotheses tests are presented in Table \ref{table:WilcoxTestsH2H3}.
For authors in $A_{\nearrow}$ we observe $p$-values much lower than the $0.01$ significance threshold of the alternative hypothesis testing (\textbf{H2}).
We hence find evidence that authors in $A_{\nearrow}$ experience a significant increase in $k$-core, betweenness and degree centrality.
Reversely, results for authors in $A_{\searrow}$ suggest a significant drop in $k$-core, eigenvector and degree centrality.
Based on these results we cannot reject hypothesis \textbf{H2} and \textbf{H3}, indicating that there is significant influence of author citation success on her future coauthorship network centrality.

\begin{figure}
  \centering
\subfigure[2002]{
  \includegraphics[width=0.4\textwidth]{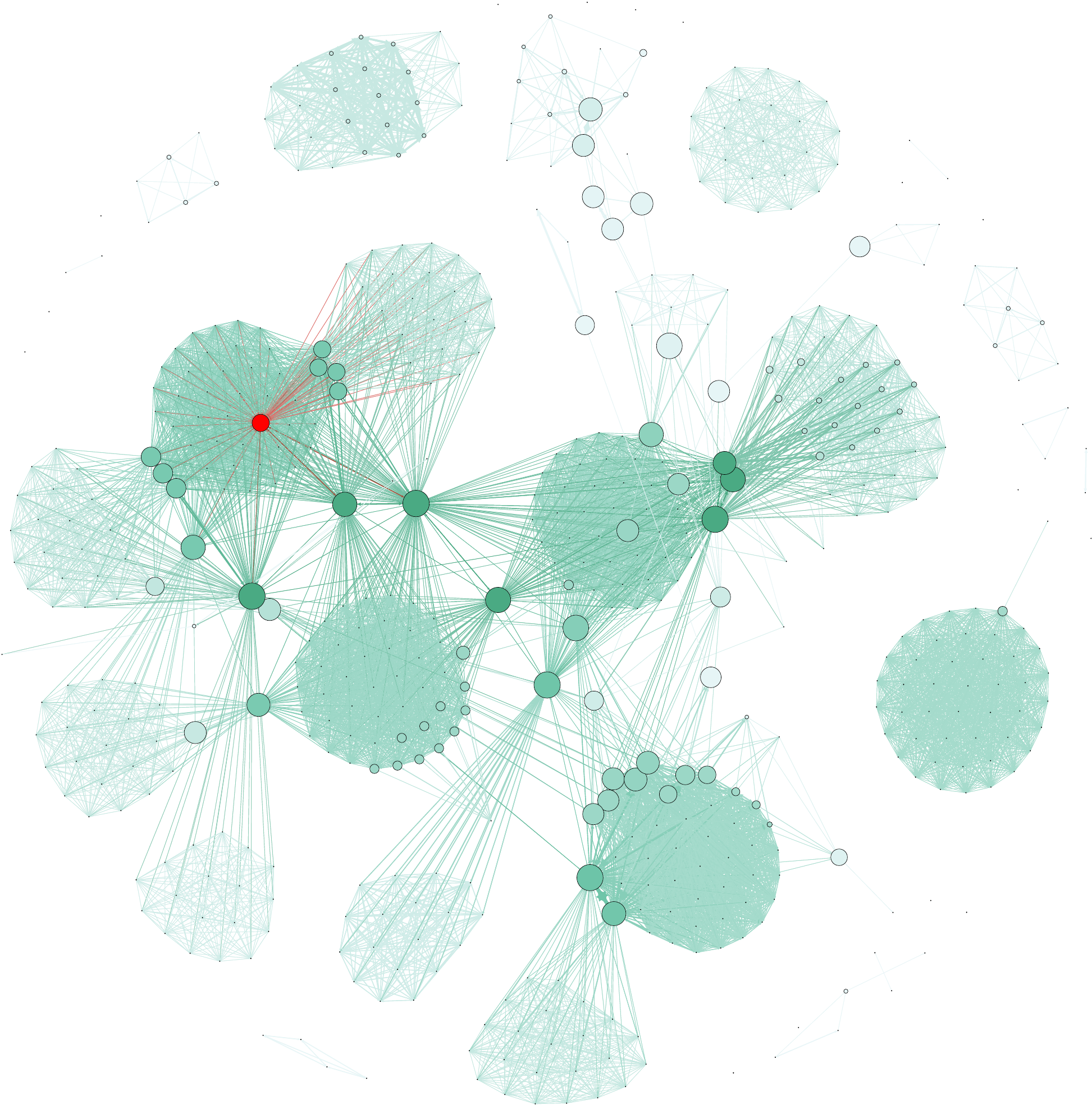}
}
\subfigure[2007]{
  \includegraphics[width=0.4\textwidth]{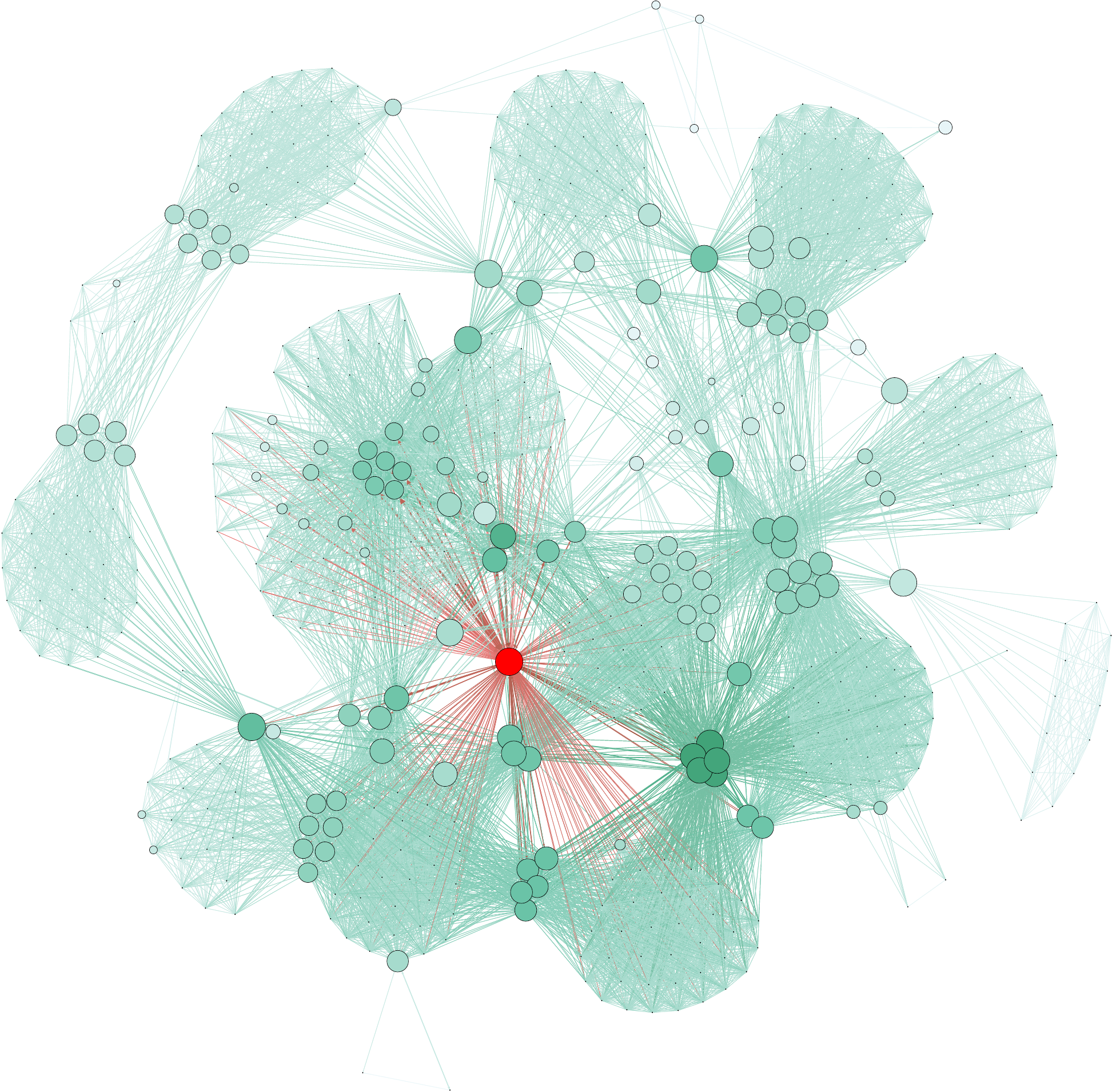}
}
  \caption{Illustration of correlation between citation success and centrality in the coauthorship network. Color intensity of the nodes is scaled according to their degree centrality and size of nodes is scaled according to their betweenness centrality.}
\label{fig:centrality_success}
\end{figure}

As an illustration of citation and coauthorship dynamics, Figure~\ref{fig:centrality_success} shows part of the coauthorship network.
Color intensity of the nodes is scaled to their degree centrality, while node size is scaled to their betweenness centrality.
A very strong community structure is clearly visible.
Furthermore, we highlighted in red one particular author that belonged to group $A_{\nearrow}(t)$, i.e. authors who did not have a paper in $P_{\uparrow}$ in 2002, but did so in 2007.
Thus, in the considered five year span the highlighted author moved from a position in the periphery of the coauthorship network to a position in the center.
Not only the authors' degree centrality increased (see size of the node as well as joined red-colored links), but also betweenness centrality improved highly.

Note that already in 2002 the author had comparatively high betweenness and degree centrality, which --according to our previous discussion-- provided an ideal starting point for citation success in 2007.

\section{Predicting Successful Publications}
\label{sec:prediction}
In the previous sections we presented evidence for the existence of statistical dependencies between authors' coauthorship centrality and the success of their publications.
Results suggested that several coauthorship centrality metrics are indicative for citation success.
However, we did not identify one single such centrality metric, especially we did not find that the mere number of coauthors is sufficient for a paper to become highly cited.
Instead, this seems to be dependent on more than one network measure.
In this section we present a machine learning classifier to predict whether a publication will be highly cited, based on several features of the authors position in the coauthorship network.

Previous works have already attempted to predict citation success.
For example in \cite{Hirsch2007}, the predictive power of the past h-index for the future h-index of a scientist was presented.
Furthermore, in \cite{Acuna2012} additional indicators like, e.g. the length of the career or the number of articles in certain journals, have been integrated into a model to predict the future h-index of scientists.
The authors of~\cite{Newman2009} compare the number of citations an article has received at a given point in time with the expected value in a preferential attachment model for the citation network.
Deriving a z-score, the authors present a prediction of which papers will be highly cited in the future.
Recently the authors reevaluate their earlier predictions and confirm the predictive power of their approach~\cite{Newman2013}.
Whereas these three approaches attempt to predict success based on past citation dynamics, they do not investigate the underlying mechanisms that lead to citation success.
Here we address this fundamental question and try to predict citation success based merely on coauthorship network centrality of authors.
Clearly, many different factors will contribute to scientific success.
In this work, however, we focus on the social component (based on the coauthorship network) in order to highlight the influence of social, not necessarily merit-based, mechanisms on publication success.

Based on the observed relations between author centralities in the coauthorship network and the success of their publications presented in section \ref{sec:hypotheses}, in this section we investigate whether we can predict a paper's future success.
In particular, we try to predict whether a paper will be highly cited five years after its publication based on measures of author centrality in the coauthorship network.

In section~\ref{sec:hypotheses:H1} we presented insights about the statistical dependency of citation success and several social network centrality measures (see Table~\ref{table:TopAuthorIntersections}).
These results suggest that a naive Bayes predictor for citation success can already yield quite useful results, predicting whether or not a paper will be $\mathsf{toppaper}$, given ex ante knowledge about $\mathsf{topmetric}$ of the authors.
Using $k$-core centrality as a basis, we apply the following classification rule:
\begin{center}
\emph{If a paper is authored by a top $10\%$ $k$-core centrality author, then the paper will be among the top $10\%$ most cited papers five years after publication}.
\end{center}
To evaluate the goodness of this prediction, we will consider the error measures \textit{precision} and \textit{recall}\footnote{See Supplementary Material for a general definition of precision and recall}.
Observing that for $k$-core centrality in a $10\%$ success scenario it is $P(\mathsf{topmetric}|\mathsf{toppaper})=0.21\%$ as well as $P(\mathsf{toppaper}|\mathsf{topmetric})=0.22\%$ and the fact that for a naive Bayes classifier $\mathsf{recall}=P(\mathsf{topmetric}|\mathsf{toppaper})$ and $\mathsf{precision}=P(\mathsf{toppaper}|\mathsf{topmetric})$ holds, one sees that a classifier with the above rule yields $\mathsf{recall}=21\%$ and $\mathsf{precision}=22\%$.
Similarly, instead of $k$-core centrality other network measures presented in Table~\ref{table:TopAuthorIntersections} can be used as basis for the above classification rule.
As earlier works have tried to predict the success of papers based on the number of coauthors~\cite{HsuHua}, using degree centrality as basis for the above classification rule directly extends these attempts, yielding $\mathsf{recall}=20\%$ and $\mathsf{precision}=20\%$.
Note, however, that degree centrality accumulates all coauthorships that have been established within the two-year sliding window of our analysis, not just the coauthorships of the paper under consideration.

We now ask whether a multi-dimensional naive Bayes classifier can improve this single metric classification result.
Taking into account the intersection of all considered centrality metrics, we consider the following classification rule:
\begin{center}
\emph{If a paper is authored by an author with a top $10\%$ betweenness centrality, degree centrality, $k$-core centrality and eigenvector centrality, then the paper will be among the top $10\%$ most cited papers five years after publication}.
\end{center}
Using this classifier, we achieve even better classification of $\mathsf{precision}=0.36\%$, however diminishing recall to $\mathsf{recall}=0.15\%$.
Whereas these results already show that a naive Bayes classifier can yield interesting insights, in the following we will present a more sophisticated Machine Learning approach, taking multiple network centrality features into account and improving classification errors.

We first construct a feature vector for every publication as follows.
For each publication appearing in year \textit{t}, we extract all coauthors and compute the maximum and minimum of their centralities in the coauthorship network constructed based on the time window [\textit{t}-2,\textit{t}].
Then, for each publication we build a feature vector with $10$ features containing the maximum and minimum of the centrality metrics considered earlier (\emph{degree}, \emph{eigenvector}, \emph{betweenness} and \emph{$k$-core}), as well as the number of coauthors and the cumulative number of authors a paper has referenced.
We then classify all publications regarding whether they fall in $P_{\uparrow}$ or $P_{\downarrow}$ according to the aforementioned publication classes, with $P_{\uparrow}$ defined as the set of the top $10\%$ cited publications and $P_{\downarrow}$ as the remaining $90\%$.

The classification is done using a Random Forest classifier \cite{Breiman2001}, extending the concept of classification trees~\footnote{We use the R package \emph{randomForest}, available at http://cran.r-project.org/web/packages/randomForest/}.
In general, the Random Forest is known to yield accurate classifications for data with a large number of features \cite{Breiman2001}.
Furthermore, it is a highly scalable classification algorithm, eliminating the need for separate cross validation and error estimation, as these procedures are part of the internal classification routine.~\footnote{For details on the procedure and the error estimates we refer to the Supplementary Material.}

Table \ref{table:PredictionValuesTop10} summarizes precision, recall, and F-score of the resulting classification.
\begin{table}
\centering
\begin{tabular}{|c|c|c|c|}
  \hline
	Nr.Publications & Precision   & Recall & F-Score \\
	\hline
	36000 	& $60 \%$         & $18 \%$     & $28 \%$  \\
\hline
\end{tabular}
\caption{Error estimates of the Random Forest classifier to predict success of papers.}
\label{table:PredictionValuesTop10}
\end{table}
Comparing this result with the expectation from a random guess, which will correctly pick one of the top $10 \%$ publications only in $10 \%$ of the cases, the achieved precision of $60 \%$ is striking.
In particular, by only considering positional features of authors in the coauthorship network, we are able to achieve \textit{an increase of factor six in predictive power} compared to a random guess.
Also, we obtain a \emph{recall} value of $18 \%$, meaning that our classifier correctly identified about one fifth of all of the top $10 \%$ papers in a given research field.
As a random guess would yield a recall of $10\%$, the Random Forest classifier \textit{improves recall by $80\%$}. 

This result allows for two conclusions:
First, the fact that a high-dimensional random forest classifier performs better than a naive Bayes classifier, makes clear that social influence on scientific success cannot be measured by a single value.
Second, and most importantly, that by \textit{solely considering metrics of social influence}, such a classifier is able to predict scientific success with high precision.

\section{Discussion and Conclusions}
\label{sec:conclusion}

Using a data set on more than \num{100000} scholarly publications authored by more than \num{160000} authors in the field of computer science, in this article we studied the relation between the centrality of authors in the coauthorship network and the future success of their publications.
Clearly, there are certain limitations to our approach, which we discuss in the following. 

First of all, any data-driven study of social behavior in general and citation behavior in particular is limited by the completeness and correctness of the used data set.
The fact that name ambiguities are automatically resolved by the Microsoft Academic Search (MSAS) database by sophisticated and validated disambiguation heuristics is a clear advantage over simpler heuristics that have been used in similar studies. 

In order to rule out effects that are due to different citation patterns in different disciplines, we limited our study to computer science, for which we expect the coverage of MSAS to be particular good.
While this limits the generalization of our results to other fields, our work nevertheless represents -- to the best of our knowledge -- the first large-scale case study of social factors in citation practices.
As publication practises seem to vary widely across disciplines, it will be interesting to investigate whether our results hold for other research communities as well.

Clearly, any study that tries to evaluate the \emph{importance} or \emph{centrality} of actors in a social network needs to be concerned about the choice of suitable centrality measures. 
In order to not overemphasize one particular -- out of the many -- dimensions of centrality in networks, we chose to use \emph{complementary centrality measures} that capture different aspects of importance at the same time.
The results of our prediction highlight that the combination of different measures is crucial -- making clear that visibility and social influence are more complicated to capture than by a single centrality measures.

Finally, one may argue that our observation that authors with high centrality are cited more often is not a statement of a \emph{direct causal relation} between centrality and citation numbers.
After all, both centrality and citations could be secondary effects of, for instance, the scientific excellence of a particular researcher, which then translates into becoming central and highly cited at the same time.
Clearly, we neither can -- nor do we want -- to rule out such possible explanations for our statistical findings.
However, considering our finding of strong statistical dependence between social centrality and citation success, one could provocatively state the following: if citation-based measures were to be good proxies for scientific success, so should then be measures of centrality in the social network.
We assume that not many researchers would approve having their work evaluated by means of such measures.
We hence think that our findings are an important contribution to the ongoing debate about the meaningfulness and use of citation-based measures, as well as a better understanding of citation dynamics in general.

In summary, the contributions of our work are threefold:
\begin{enumerate}
\item  We provide the, to the best of our knowledge, first large-scale study that analyses relations between the position of researchers in scientific collaboration networks and citation dynamics, using a set of complementary network-based centrality measures.
A specific feature of our method is that we study \emph{time-evolving} collaboration networks and citation numbers, thus allowing us to investigate possible mechanisms of social influence at a microscopic scale.
\item We show that -- at least for the measures of centrality investigated in this paper -- there is no \emph{single} notion of centrality in social networks that could accurately predict the future citation success of an author. 
We expect this finding to be of interest for any general attempt to predict the success of actors based on their centrality in social networks.
\item Using modern machine learning techniques, we present a supervised classification method based on a Random Forest classifier, using a multidimensional feature vector of collaboration network centrality metrics.
We show that this method allows for a remarkably precise prediction of the future citation success of a paper, solely based on the social embedding of its authors. 
With this, our method provides a clear indication for a strong statistical dependence between author centrality and citation success.
\end{enumerate}

In conclusion, we provided evidence for a strong relation between the position of authors in scientific collaboration networks and their future success in terms of citations.
We would like to emphasize that by this we \emph{do not} want to join in the line of -- sometimes remarkably uncritical -- proponents of citation-based evaluation techniques.
Instead, we hope to contribute to the discussion about the manifold influencing factors of citation measures and their explanatory power concerning scientific success.
Especially, we \emph{do not} see our contribution in the development of automated success prediction techniques, whose widespread adoption could possibly have devastating effects on the general scientific culture and attitude.
Highlighting social influence mechanisms, we rather hope that our work contributes to a better understanding of the multi-faceted, complex nature of citations, which should be a prerequisite for any reasonable application of citation-based measures.
\section{Acknowledgement}
\label{sec:acknowledgement}
EM, IS and FS acknowledge funding by the Swiss National Science Foundation, grant no. \texttt{CR31I1\_140644/1}.
AG acknowledges funding by the EU FET project \textsc{Multiplex} \texttt{317532}.
We especially thank Microsoft Research for granting unrestricted access to the Microsoft Academic Search service.
\bibliographystyle{abbrv}
\bibliography{paper}

\clearpage
\section*{Supplementary Material}

\subsection*{Centrality Metrics}

There are many network metrics that can be used for social network analysis~\cite{WassermanFaust, NewmanBook}.
Here we are quickly going to review the metrics we have been using in this work and their interpretation in coauthorship networks.

\emph{Degree Centrality}
The degree centrality of a node is its number of first-order neighbors, i.e. the number of nodes this nodes connects to via one link.
In a directed network, this measure is divided into an in-degree centrality and an out-degree centrality.
Since the here considered coauthorship network is undirected, the degree centrality is simply the number of its direct neighbors.
Degree centrality is a local measure, as it does not depend on any global network properties other than the number of its neighbors.
In the coauthorship network the degree centrality of a node is its number of coauthors.

\emph{Eigenvector Centrality}
The eigenvector centrality of a node is a global centrality measure, as non-local changes in the network can alter the node's eigenvector centrality.
In short, a node has high eigenvector centrality if it is connected to other nodes with high eigenvector centrality.
As such, this centrality measure goes beyond degree centrality as a mere measure of quantity (the number of neighbors) in that it introduces a notion of inheritance of importance.
Used often, especially in its variant PageRank, eigenvector centrality of node $v$ is the $v$th component of the Perron-Frobenius-eigenvector of the network's adjacency matrix.
In the coauthorship network, eigenvector centrality has a meaning of importance, if one assumes that an author is more important if she coauthors papers with other authors of high importance.

\emph{Betweenness Centrality}
The Betweenness centrality is another often used global centrality measure.
A node has high betweenness centrality if it lies on many shortest paths of the network.
Hence, this centrality measure is a measure of importance in terms of network flows.
If a node with high betweenness centrality would be removed, a lot of network flows would become less efficient, as the average length of shortest paths will increase.
In the coauthorship network, a node with high betweenness centrality could be interpreted as a node with high importance for ``fast knowledge transfer'', as this person lies on many shortest paths connecting authors and their research.

\emph{K-Core Centrality}
The $k$-core centrality is a global centrality measure thought to measure the ``coreness'' of a node, i.e. how deep it is embedded in the network.
A node has $k$-core centrality $k$ if, when consecutively removing nodes that have degree $1$, $2$, ...$k-1$ from the network, this node has not been removed, but will be removed in a next step when nodes with degree $k$ are removed.
$k$-core centrality is somewhat similar to eigenvector centrality, as a node must have neighbors with high $k$-core in order to have high-core itself.
Different from eigenvector centrality, $k$-core centrality is not additive. 
Hence compensating a low number of high $k$-core neighbors with a high number of low $k$-core neighbors does not guarantee the node to have high $k$-core.
In the coauthorship network a node has a high $k$-core, if it is connected to many nodes that have high $k$-core themselves.

\subsection*{Correlations Between Citation Numbers and Centrality Metrics}

In section \textit{Effects of Author Centrality on Citation Success} of the main manuscript, we argue that citation numbers of an article, five years after its publication, are not Pearson- and Spearman-correlated with social network centrality metrics of its authors. 
Table \ref{table:correlations} summarizes the Pearson and Spearman correlation coefficients for the considered metrics.
None of these results allows to conclude any significant correlation.
\begin{table}[h]
\centering
\begin{tabular}{
|
c|
S[table-format = 1.2]|
S[table-number-alignment=center, table-format = 1.2]
|
}
      \hline
	& {Pearson $r$} 	& {Spearman $\rho$}  \\
	\hline
	\textbf{$k$-core} & 0.05 & 0.15\\
		\hline
		
	\textbf{Eigenvector} & 0.01 & 0.07 \\
		\hline
		
	\textbf{Betweenness} & 0.13 & 0.15 \\
    	\hline
    	
	\textbf{Degree} & 0.05 & 0.16 \\
    	\hline

\end{tabular}
\caption {Pearson and Spearman coefficients measuring correlations between citation numbers of a paper (five years after publication) and coauthorship network centrality of its authors.}
\label{table:correlations}
\end{table}
\subsubsection*{Precision and Recall}
In Machine Learning it is standard practice to assess the goodness of a classifier using the quantities \emph{precision} and \emph{recall}~\cite{goutte2005probabilistic}.

\emph{Precision} is defined as
\begin{equation}
Precision = \frac{TruePositives}{TruePositives+FalsePositives}.
\end{equation}

It hence is equal to the fraction of correctly predicted instances compared to all predicted instances.
As such, precision quantifies how reliable the predicted results are.
However, it does not make any statement about how many relevant results the predictor returns.
For example, a simple predictor could be to always return one element from which it is known ex ante, that it is a true prediction.
In this scenario precision would be $100\%$, however the sensitivity might be poor as there might be more than one relevant element.
This last point is quantified using \emph{recall}.

\emph{Recall} (or \emph{sensitivity}) is defined as
\begin{equation}
Recall = \frac{TruePositives}{Positives}.
\end{equation}

In a case of 100 samples, 1 positive and 99 negative, a classifier that solely returns the one positive element has $recall=1$ as well as $precision=1$.
However, a classifier that returns all 100 elements still has $recall=1$, although precision will be $precision=1/99$.

Precision and recall are usually presented jointly to assess the goodness of a classifier. Sometimes they are combined to the so called \emph{F-score}:
\begin{equation}
F=2 \cdot \frac{precision \cdot recall}{precision + recall}
\end{equation}

F-score value $F=1$ indicates best possible classifier goodness, $F=0$ indicates worst.
\subsection*{Random Forest Prediction}

In section four \textit{Predicting Successful Publications} of the main manuscript, we use a Random Forests classifier~\cite{Breiman2001} to predict success of publications based on coauthorship centrality of its authors.
A Random Forest classifier fits a number of classification trees (a so called forest) on bootstrap samples of the data set with subsequent averaging over all individual classification trees to improve predictive accuracy.
Random Forests perform well on classification tasks involving a large number of features.
We will quickly outline the high-level procedure, for details please see~\cite{Breiman2001}.

\begin{enumerate}

\item A bootstrap sample from the minority class, and a sample with the same number of cases from the majority class are drawn, with replacement, for each iteration of the Random Forest classification.

\item For each iteration, a classification tree from the data is grown, without pruning. Assuming that there are \textit{M} variables in the data set, a number \textit{m<<M} is chosen such that at each node, \textit{m} variables are selected at random out of the \textit{M} and the best split on these \textit{m} is used to split the node. The best split is discovered using the CART algorithm, where, at each node, instead of searching through all variables for the optimal split, only \textit{m} variables are searched through.

\item The two steps above are repeated depending on how many trees are desired to be grown. After a bootstrap sample of the data is put down each tree, each tree votes for what it computes as the true classes for each case in that sample set. The final classification for each case is the one having most votes among the individual trees. 
\end{enumerate}

In this work, we utilize the R-package \textit{randomForest}\footnote{http://cran.r-project.org/web/packages/randomForest/} to perform the classification.

\subsubsection*{Out-of-bag (OOB) Error Estimate}

Random Forests do not require a separate test set to cross-validate the classification and to estimate the classification error rate. 
Instead, error estimation can be done during run time of the algorithm. 
Each tree is constructed using a different bootstrap sample from the original data where about one-third of the cases from the sample are not used in the construction. 
These are called the \textit{Out of bag (OOB)} cases. 
These \textit{OOB} cases are later classified using the constructed decision trees.
At the end of the classification, an error estimate is computed by averaging the proportion of times where the class that got the most votes is not equal to the true class over all OOB cases.

\subsubsection*{Importance and Significance of Features}

In addition to OBB error estimation, Random Forest classifiers quantify the significance and importance of features. 

\begin{enumerate}
  \item For every tree in the forest which uses a sample with \textit{m} features, the number of votes for the correct classification of the \textit{OOB} cases are calculated.
  \item Then the values of feature \textit{k} in the \textit{OOB} cases are permuted, and the correctly classified cases from this permutation is also calculated.
  \item The average of the difference of counts from step 1 and step 2 over all trees is the importance score of feature \textit{k}.
  \item If the values of this score are independent across individual trees, the standard error can be computed by dividing the raw score by its standard error to get a \textit{z-score} and assigning a significance level to the feature.
  
\end{enumerate}

Taking samples of three variables at a time ( \textit{m} = 3 ) for each split, we classify the publications. Using the internal variable importance ranks, we are able to detect which variables hold the most predictive power for classifying true positives.

\end{document}